\documentstyle[12pt,epsf]{article}
\newcommand{\OO}{{\cal O}}

\newcommand{\LL}{{\cal L}}

\newcommand{\half}{{1\over 2}}
\newcommand{\wt}{\widetilde}
\newcommand{\wh}{\widehat}

\newcommand{\wb}{\bar}

\newcommand{\be}{\begin{equation}}
\newcommand{\ee}{\end{equation}}
\newcommand{\ben}{\begin{eqnarray}\displaystyle}
\newcommand{\een}{\end{eqnarray}}
\newcommand{\refb}[1]{(\ref{#1})}
\newcommand{\p}{\partial}
\newcommand{\sectiono}[1]{\section{#1}\setcounter{equation}{0}}

\begin{document}

\begin{flushright}
hep-th/0003278\\
MRI-PHY/P20000310
\end{flushright}

\vskip 3.5cm

\centerline{\large \bf Tachyon Condensation and Brane Descent Relations}

\medskip
\centerline{\large \bf  in p-adic String Theory}

\vspace*{6.0ex}

\centerline{\large \rm Debashis Ghoshal\footnote{E-mail:
ghoshal@mri.ernet.in} and Ashoke Sen\footnote{E-mail: 
ashoke.sen@cern.ch, sen@mri.ernet.in}}

\vspace*{1.5ex}

\centerline{\large \it Mehta Research Institute of Mathematics}
 \centerline{\large \it and Mathematical Physics}

\centerline{\large \it  Chhatnag Road, Jhusi,
Allahabad 211019, INDIA}

\vspace*{4.5ex}

\centerline {\bf Abstract}

It has been conjectured that an extremum of the tachyon potential of 
a bosonic D-brane represents the vacuum without any D-brane, and that 
various tachyonic lump solutions represent D-branes of lower dimension.
We show that the tree level effective action of $p$-adic string theory,
the expression for which is known exactly, provides an explicit 
realisation of these conjectures.

\vfill \eject

\tableofcontents

\baselineskip=18pt

\sectiono{Introduction and Summary} \label{s1}

The world-volume theory on the D-brane of a bosonic string theory contains
a tachyonic mode. It has been conjectured that the tachyon potential has
a non-trivial extremum where the potential energy of the tachyon
exactly cancels the tension of the D-brane, and that this
configuration represents the closed string vacuum without any
D-brane\cite{9902105}. It
has been further conjectured that various tachyonic lump solutions on the
D-brane world-volume represent D-branes of lower
dimensions\cite{RECK,9902105}. These
conjectures and their generalisations to superstring
theories\cite{9805019}--\cite{9812135} have
been tested by various
methods\cite{RECK,9812031},\cite{9808141}--\cite{0003110}.
However, since the exact
effective
action for the tachyon
field is not known, there is no direct proof of these
conjectures.

In this paper we point out that in the $p$-adic string theory
introduced in \cite{FROL}--\cite{HLSP} (see \cite{BRFR} for a review) 
we can explicitly check these conjectures. It should be emphasised
that although the $p$-adic `string' is an exotic object, the 
spacetime it describes is the familiar one\footnote{A different type 
of $p$-adic string was considered in \cite{Volovich}.} . 
In the $p$-adic open string theory, which in modern language can 
be regarded as the world-volume theory of a space-filling 
D-brane, the {\em exact} classical action of the tachyon field and 
various solutions of the equations of motion are known\cite{padic}. 
Among the known non-trivial solutions is a translationally invariant 
solution with the property that it is a local minimum of the potential, 
and that the propagator of the tachyon field describing fluctuations 
around this background has no physical pole. Thus this configuration 
has no physical open string excitations, and is naturally identified 
with the vacuum without a D-brane. The exact tachyon equation of 
motion of the $p$-adic string theory also has classical lump 
solutions for all codimension $\ge 1$, 
which approach the vacuum solution far away from the core of the
soliton. If the original open string 
theory is formulated in  $(d-1,1)$ dimensional space-time\footnote{
There is as yet no compelling reason for a critical dimension in
$p$-adic string theory, but the so called adelic 
formula\cite{FRWI,BRFR} relating the product of 
four tachyon amplitudes in $p$-adic strings for all primes $p$ to
that in the bosonic string suggests that they all have the same
critical dimension $d=26$.}, then such a lump solution of codimension 
$(d-q-1)$ describes a solitonic $q$-brane. We show that the world-volume 
theory on the solitonic $q$-brane agrees with the expected world-volume 
theory on a Dirichlet $q$-brane in the $p$-adic string theory, 
to the extent that we can compare them with 
the present knowledge. This provides strong evidence that these lump 
solutions can be identified as lower dimensional D-branes.

The paper is organised as follows. In section \ref{s2} we summarise 
the exact effective action of the tachyon in the $p$-adic string theory, 
the known solutions of the equation of motion derived from the 
action and their properties. In section \ref{s3} we analyse the 
world-volume theory of the solitonic $q$-brane, and in section
\ref{s4} we compare this with the world-volume theory of a Dirichlet 
$q$-brane. Section \ref{s5} contains some comments on further 
extension of this work, and ends with speculation on its possible 
application to the study 
of tachyon condensation in ordinary bosonic string theory.

\sectiono{Solitonic $q$-branes of $p$-adic string theory} \label{s2}

In ref.\cite{FROL} $p$-adic string theory was defined as follows. 
Consider the expressions for various amplitudes in ordinary bosonic open 
string theory, written as integrals over the boundary of the world-sheet
which is the real line {\bf R}. Now replace the integrals over {\bf
R} by integrals over the $p$-adic field ${\bf Q}_p$ with appropriate
measure, and the norms of the functions in the integrand by the
$p$-adic norms. These rules were subsequently derived from a local
action defined on the ``world-sheet'' of the $p$-adic 
string\cite{ZABRODIN,BRFR}. Using $p$-adic analysis, it is possible
to compute $N$ tachyon amplitudes at tree-level for all $N\ge 3$. 

\begin{figure}[!ht]
\leavevmode
\begin{center}
\epsfbox{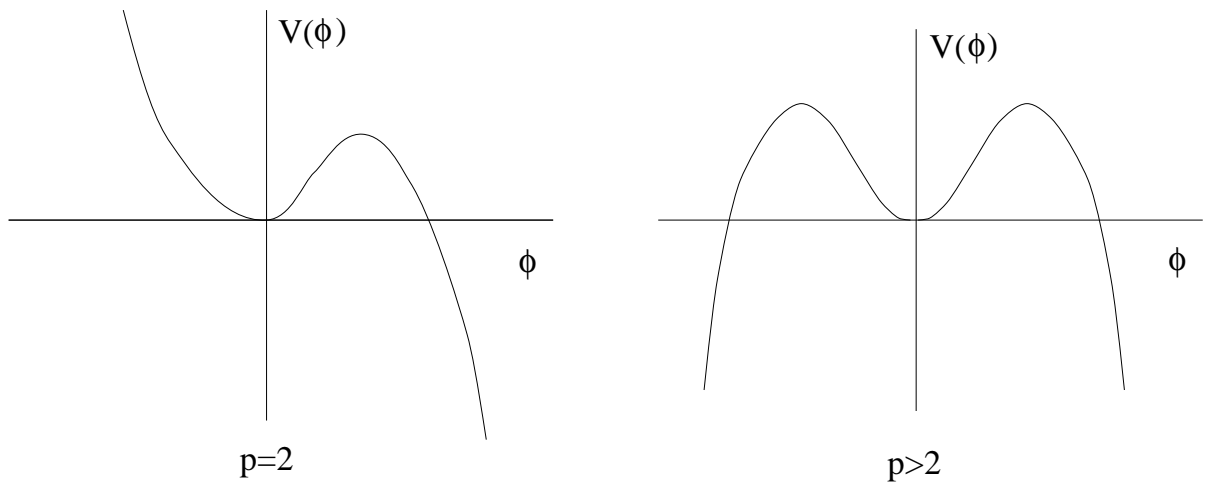}
\caption{The effective tachyon potential for the $p$-adic string.}
\label{f2}
\end{center}
\end{figure}

This leads to an {\em exact} action for the open string tachyon in 
$d$ dimensional $p$-adic string theory. This action is given in
ref.\cite{padic}
\ben \label{e1}
S &=& \int d^d x \LL \nonumber \\ 
&=& {1\over g^2} {p^2 \over p-1} \int d^d x \left[ -{1\over 2} 
\phi p^{-\half\Box}\phi + {1\over p+1} \phi^{p+1} \right]\, ,
\een
where $\Box$ denotes the $d$ dimensional Laplacian, $\phi$ is the 
tachyon field (after a rescaling and a shift), $g$ is the open 
string coupling constant, and $p$ is an arbitrary prime number.
We are using metric with signature 
$(-,+,+\ldots +)$. If we denote by $(2\pi\alpha'_p)^{-1}$ the `tension
of the $p$-adic string' as defined in ref.\cite{ZABRODIN}, then our
choice of units correspond to\cite{ZABRODIN}
\be \label{etension}
\alpha'_p = {\ln p\over 2\pi}\, .
\ee
We have added a
constant term to the Lagrangian 
density $\LL$ so that it vanishes at $\phi=0$. Fig.\ref{f2} shows the
qualitative features of the tachyon potential for different values
of $p$. 

The equation of motion derived from this action is,
\be \label{e1b}
p^{-\half\Box} \phi = \phi^p\, .
\ee

Different known solutions of this equation are as follows\cite{padic}:
\begin{itemize}
\item The configuration $\phi=1$ is the original vacuum around 
which we quantised the string\footnote{For $p\ne 2$, there is also 
an equivalent solution corresponding to $\phi=-1$. Since the action
is symmetric under $\phi\to -\phi$, we shall restrict our analysis to
solutions with positive $\phi$.}. We shall call this the 
D-$(d-1)$-brane solution. The energy density associated with this 
configuration, which can be identified as the tension $T_{d-1}$ of 
the D-$(d-1)$-brane configuration, is given by 
\be \label{e1a}
T_{d-1} = - \LL(\phi=1) = {1\over 2 g^2} {p^2 \over p+1}\, .
\ee

\item The configuration $\phi=0$ denotes a configuration around which
there is no perturbative physical excitation. We shall identify this
with the vacuum configuration. By definition we have taken
the energy density of this vacuum to be zero.

\item The configuration:
\be \label{e2}
\phi(x) = f(x^{q+1}) f (x^{q+2}) \cdots f(x^{d-1})\equiv
F^{(d-q-1)}(x^{q+1},\ldots,x^{d-1})\, ,
\ee
with
\be \label{e3}
f(\eta) \equiv p^{1\over 2(p-1)} \exp\left(-{1\over 2}\, 
{p-1\over p \ln p} \, \eta^2\right) \, ,
\ee
describes a soliton solution with energy density localised 
around the hyperplane $x^{q+1} = \cdots = x^{d-1}=0$. This follows 
from the identity:
\be \label{es1}
p^{-\half \p_\eta^2} f(\eta) = \left(f(\eta)\right)^p\, .
\ee
We shall call \refb{e2}, with $f$ as in \refb{e3}, 
the solitonic $q$-brane solution. Let us denote by
$x_\perp=(x^{q+1},\ldots, x^{d-1})$ the coordinates transverse to the 
brane and by $x_\parallel=(x^0,\ldots,x^q)$ those tangential to it. 
The energy density per unit $q$-volume of this brane, which can 
be identified as its tension $T_q$, is given by
\be \label{e4}
T_q = - \int d^{d-q-1} x_\perp\; \LL(\phi = F^{(d-q-1)}(x_\perp))
= {1\over 2g_q^2} \, {p^2 \over p+1}\, ,
\ee
where,
\be \label{e9}
g_q = g \left[{p^2 - 1\over 
2 \pi\, p^{2p/(2p-1)}\ln p}\right]^{(d-q-1)/4}\, .
\ee
\end{itemize}
{}From eqs.\refb{e1a},\refb{e4} and \refb{e9} we see that the ratio
of the tension of a $q$-brane to a $(q-1)$-brane is
\be \label{e5}
{T_q\over T_{q-1}} = \left[ { 2\pi\, p^{2p\over p-1}\ln p \over
p^2 -1}\right]^{-{1\over 2}} =
{\sqrt{p^2-1}\over p^{p\over p-1}}\;{1\over 2\pi\sqrt{\alpha_p'}} \, .
\ee
In the above equation we have used dimensional analysis and 
\refb{etension} to restore factors of $\alpha_p'$. 
Note that the ratio \refb{e5} is independent of $q$.
This is a feature of the D-branes in ordinary 
bosonic string theory, and suggests that the solitonic $q$-branes of 
$p$-adic string theory should have interpretation as 
D-branes. This also suggests that the self-dual radius $R_{sd}$ of the 
$p$-adic string theory, where the tension $2\pi R_{sd}T_q$ of a wrapped
$q$-brane is
equal to the tension $T_{q-1}$ of a $(q-1)$-brane, is given by 
\be \label{eselfdual}
R_{sd} = { p^{p/(p-1)}\over \sqrt{p^2-1}}\, \sqrt{\alpha'_p}\, .
\ee
Note that as $p\to\infty$, this approaches the formula for the self-dual
radius in ordinary bosonic string theory.

\sectiono{World-volume theory on the solitonic $q$-branes} \label{s3}

Let us now consider a configuration of the type
\be \label{e6}
\phi(x) = F^{(d-q-1)}(x_\perp) \psi(x_\parallel) \, ,
\ee
with $F^{(d-q-1)}(x_\perp)$ as defined in \refb{e2},\refb{e3}.
For $\psi=1$ this describes the solitonic $q$-brane. Fluctuations of
$\psi$ around 1 denote fluctuations of $\phi$ localised on the
soliton; thus $\psi(x_\parallel)$ can be regarded as one 
of the fields on its world-volume. We shall call this 
the tachyon field on the solitonic $q$-brane world-volume\footnote{In 
the linearised approximation this tachyonic mode was discussed in
ref.\cite{FRNI}.}.
Substituting \refb{e6} into \refb{e1b} and using \refb{es1} we get
\be \label{e7}
p^{-\half\Box_\parallel} \psi = \psi^p\, ,
\ee
where $\Box_\parallel$ denotes the $(q+1)$ dimensional Laplacian 
involving the world-volume coordinates $x_\parallel$ of the $q$-brane. 
The action involving $\psi$ can be obtained by substituting \refb{e6} 
into \refb{e1}:
\ben 
S_q(\psi) &=& S\left(\phi=F^{(d-q-1)}(x_\perp)\psi(x_\parallel)\right)
\nonumber\\
&=& {1\over g_q^2} {p^2 \over p-1} \int d^{q+1} x_\parallel 
\left[ -{1\over 2} \psi p^{-\half \Box_\parallel}
\psi + {1\over p+1} \psi^{p+1} \right]\, ,\label{e8}
\een
where $g_q$ has been defined in eqn.\refb{e9}. 

Note that a solution of \refb{e7} gives an {\em exact} solution of the 
full equation of motion \refb{e1b}. Thus eq.\refb{e7} describes 
the dynamics of the mode $\psi$ on the $q$-brane world-volume exactly. 
This does not mean that there are no other modes on the $q$-brane 
world-volume; rather what this implies is that it is possible to obtain 
a consistent truncation of the world-volume theory of the
$q$-brane by setting 
all the modes except $\psi$ to zero. In terms of scattering amplitudes 
this means that the tree level S-matrix on the $q$-brane world-volume, 
involving {\em only} external tachyon states, can be calculated 
exactly from the action \refb{e8}.

Of the various other (infinite number of) modes living on the
$q$-brane world-volume are the $(d-q-1)$ massless modes $\xi^i$
associated with translations of the brane in the $(d-q-1)$ directions 
$x_\perp$ transverse to the brane. Inclusion of these modes correspond 
to deformation of $\phi$ of the form
\be \label{e11}
\phi(x) = F^{(d-q-1)}(x_\perp)\, \psi(x_\parallel) + \p_{x_\perp^i}\! 
F^{(d-q-1)}(x_\perp)\, \xi^i(x_\parallel) + \cdots\, .
\ee
Substituting this in eq.\refb{e1b}, and comparing the coefficients of 
the independent functions $\left(F(x_\perp)\right)^p$ and 
$\left(F(x_\perp)\right)^{p-1}\p_{x_\perp^i}\! F(x_\perp)$ on both 
sides, we get the following equations of motion:
\ben \label{e12}
p^{-\half\Box_\parallel} \psi &=& \psi^p + \OO(\xi^2)\, \nonumber \\
p^{-\half\Box_\parallel} \xi^i &=& \psi^{p-1}\xi^i + \OO(\xi^2)\, .
\een
The above equations can be derived from the effective action:
\ben \label{e13}
S_q(\psi,\xi^i) 
&=& {1\over g_q^2} {p^2 \over p-1} \int d^{q+1} x_\parallel \bigg[ 
-{1\over 2} \psi p^{-\half \Box_\parallel}
\psi + {1\over p+1} \psi^{p+1}\nonumber \\
&& - C \left\{ {1\over 2} \xi^i
p^{-\half \Box_\parallel} \xi^i - {1\over 2} \psi^{p-1} 
\xi^i \xi^i\right\} + \OO(\xi^3)\bigg]\, .
\een
$C$ is a normalisation constant whose value is not important to this
order, as it
can be changed by rescaling $\xi^i$.

$\psi=1$ corresponds to the solitonic $q$-brane solution. For
computing amplitudes involving the world-volume fields on the 
$q$-brane, we define the shifted field $\sigma$ and rescaled fields 
$\chi^i$ through the relation
\be \label{e14}
\psi = 1 + {g_q \sigma \over p}\, , \qquad 
\xi^i = {g_q\over \sqrt{p \, C}}\, \chi^i\, ,
\ee
and expand the action \refb{e13} in powers of $\sigma$ and $\chi^i$. 
This gives 
\ben \label{e15}
S_q &=& {p \over p-1} \int d^{q+1}x_\parallel \Bigg[ -{1\over 2} 
\sigma p^{-\half \Box_\parallel -1}
\sigma + {1 \over g_q^2} \, {p\over p+1} \left(1 + {g_q \sigma\over p}
\right)^{p+1} - {\sigma\over g_q} - {p \over 2 g_q^2} \nonumber \\
&&\qquad\qquad\qquad\quad - {1\over 2} \chi^i p^{-\half \Box_\parallel}
\chi^i + {1\over 2} \left(1 + {g_q \sigma \over p}\right)^{p-1} 
\chi^i \chi^i + \OO(\chi^3)\Bigg]\, .
\een
There is no linear term in $\sigma$ in action \refb{e15} reflecting
the fact that $\sigma=0$ is a solution of the equation of motion.
The momentum space $\sigma$ and the $\chi^i$ propagators computed from
this action are
given by:
\ben \label{e16}
\Delta_{\sigma\sigma}(k) &=& -\, i\; {p-1\over p} {1 \over p^{\half k^2 -1}
-1}\, ,
\nonumber \\
\Delta_{\chi^i\chi^j}(k) &=& -\, i\; {p-1\over p} {1 \over p^{\half k^2}
-1} \delta_{ij}\, .
\een
The residues at the poles in the $\sigma$ and the $\chi^i$ propagators 
(at $k^2=2$ and $k^2=0$ respectively) have the same values. This will help 
us compare the amplitudes involving $\sigma$'s and $\chi^i$'s in the 
external legs.

\sectiono{Comparison with the world-volume theory on a Dirichlet
$q$-brane} \label{s4}

We shall now compare the world-volume action on the solitonic $q$-brane
with that on the D-$q$-brane. We are immediately
faced with the question whether it is possible to have {\em Dirichlet}
branes in $p$-adic string theory. Fortunately, a `world-sheet' approach
to $p$-adic string has been developed in ref.\cite{ZABRODIN}. 
According to these authors, this `world-sheet' is a so called 
Bruhat-Tits tree --- a Bethe lattice with $p+1$ nearest neighbours --- 
the `boundary' of which is
the field of $p$-adic numbers ${\bf Q}_p$. The generalisation of the 
Polyakov action is the lattice discretisation of the action for free 
scalar fields corresponding to the target space coordinates. Now
one can either choose Neumann or Dirichlet boundary conditions as in
the case of ordinary strings. It was shown in \cite{ZABRODIN} that 
Neumann boundary conditions leads to the tachyon amplitudes 
postulated in \cite{FROL,padic}. 

While this may be the proper way to define D-branes in $p$-adic string
theory, we shall content ourselves with the continuation of the
relevant formul\ae\ from ordinary bosonic string theory. Thus, for our 
purposes the world-volume theory of a $p$-adic D-$q$-brane is defined 
by taking the expressions for various amplitudes for an ordinary 
D-$q$-brane, written as integrals over world-sheet coordinates of the 
appropriate vertex operators, 
and then replacing the integrals over real line by integrals over the
$p$-adic field, with all the norms appearing in the integrand replaced 
by $p$-adic norms as in ref.\cite{padic}. In principle one should be 
able to derive these rules from the world-sheet description in
ref.\cite{ZABRODIN}.

For amplitudes involving the external tachyons, described by the vertex
operators of the type $e^{ik\cdot X_\parallel}$ with momentum $k$ 
restricted to lie along the 
world-volume of the D-$q$-brane, the computation of the amplitude is 
identical to the one described in ref.\cite{padic}. Thus, following the
analysis there, these S-matrix elements can be obtained from an effective 
action of the form:
\be \label{e17}
\wh S_q(\psi)
= {1\over \wh g_q^2} {p^2 \over p-1} \int d^{q+1}x_\parallel \left[ 
-{1\over 2} \psi p^{-\half \Box_\parallel}\psi + {1\over p+1} 
\psi^{p+1} \right]\, ,
\ee
where the tachyon field $\psi$ is shifted so that $\psi=1$ describes the 
D-$q$-brane, and $\wh g_q$ denotes the coupling constant which 
characterises the strength of the interaction in the world-volume theory 
of the D-$q$-brane. Comparing this with \refb{e8} we see that the 
world-volume actions for the tachyon fields on the solitonic $q$-brane 
and the Dirichlet $q$-brane agree exactly if we choose:
\be \label{e18}
\wh g_q = g_q\, .
\ee
At present there is no independent derivation of $\wh g_q$ in terms 
of $g$, and hence we cannot verify eqn.\refb{e18} 
independently\footnote{This is related to the problem of computing the 
tension of the D-$q$-brane independently.}. But assuming \refb{e18} to be 
true, we have a complete agreement between the world-volume theories
involving the tachyon fields on the D-$q$-brane and the solitonic 
$q$-brane.

\begin{figure}[!ht]
\leavevmode
\begin{center}
\epsfbox{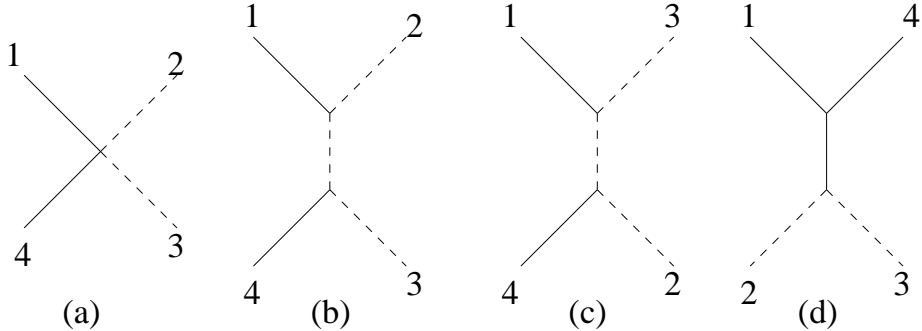}
\caption{The Feynmann diagrams contributing to the amplitude
$\langle\sigma\sigma\chi\chi\rangle$
on the solitonic $q$-brane. A dashed line denotes the $\chi$
propagator and a solid line denotes the $\sigma$ 
propagator.} \label{f1}
\end{center}
\end{figure}

Next we shall compare the amplitude
$\langle\sigma\sigma\chi^i\chi^j\rangle$ on the solitonic $q$-brane 
and the D-$q$-brane. First let us compute this on the solitonic 
$q$-brane using the action \refb{e15}. The four Feynmann diagrams 
contributing to it have been shown in
Fig.\ref{f1}. These can be easily evaluated, and the answer is:
\be \label{e19}
A_{ij}(k_1, k_2, k_3, k_4) = \delta_{ij}\, g_q^2\, \left[{p-2\over p}
+ {p-1\over
p} \, 
\left\{ {1\over p^{k_1\cdot k_2 +1} -1} + {1\over p^{k_1\cdot k_3 +1} -1} 
+ {1\over p^{k_1\cdot k_4 +1} -1}\right\}\right]\, ,
\ee
with the contribution to the four terms in the right hand side of
\refb{e19} coming from the Feynman digrams (a), (b), (c) and (d)
respectively in Fig.\ref{f1}. In deriving \refb{e19} we have used the
mass-shell conditions
\be \label{e20}
k_1^2 = k_4^2 =2, \qquad k_2^2=k_3^2 =0\, .
\ee
Let us now evaluate the same amplitude on the D-$q$-brane. The 
vertex operator associated with the mode $\chi^i$ on the D-$q$-brane 
is given by\footnote{The notation $\p X$ is
schematic, as care is needed to define the correct vertex operator
that corresponds to the analogous one for ordinary string.}
$\p X_\perp^i e^{i k\cdot X_\parallel}$. 
Inserting the $\chi^i$ and $\chi^j$ vertex operators carrying momenta 
$k_2$ and $k_3$ at 0 and 1 respectively, and the two $\sigma$ vertex 
operators carrying momenta $k_1$ and $k_4$ at $x$ and $\infty$ 
respectively, we can express the amplitude as:
\be \label{e21}
\wh A_{ij}(k_1,k_2,k_3,k_4)= \delta_{ij} \, \wh g_q^2\, \int_{{\bf Q}_p}
dx
|x|^{k_1\cdot k_2} |1-x|^{k_1\cdot k_3}\, .
\ee
Here $|~|$ denotes the $p$-adic norm and integral over $x$ is over 
the $p$-adic field. This is precisely the integral evaluated in
\cite{padic}. Using the identity
\be \label{e22}
k_1\cdot k_2 + k_1\cdot k_3 + 2 = - k_1 \cdot k_4\, ,
\ee
we can express this amplitude as
\be \label{e23}
\wh A_{ij}(k_1, k_2, k_3, k_4) = \delta_{ij}\, (\wh g_q)^2\, 
\Big[{p-2\over p} +
{p-1\over
p} \, 
\Big\{ {1\over p^{k_1\cdot k_2 +1} -1} + {1\over p^{k_1\cdot k_3 +1} -1} 
+ {1\over p^{k_1\cdot k_4 +1} -1}\Big\}\Big]\, .
\ee
This agrees precisely with eq.\refb{e19} for $\wh g_q=g_q$. 

In fact, it is possible to give a general argument showing that an
amplitude with two external $\chi$ fields and $N$ external $\sigma$
fields for arbitrary $N$,
computed from the action \refb{e15}, agrees with the corresponding
amplitude on a Dirichlet $q$-brane. To see this, let us consider 
the situation where we start with the action \refb{e1} with $g$ 
replaced by another coupling constant $\wb g$, and 
compactify\footnote{Alternatively, we can work with the  
uncompactified theory, but just examine those modes of $\phi$ which 
carry either 0 or $\pm\sqrt 2$ units of momentum in $(d-q-1)$ of 
the directions.} $(d-q-1)$ directions on circles of radii 
$1/\sqrt{2}$. Let $u^i$ denote the compact coordinates and  
$z^\mu$ the non-compact ones, and consider an expansion of the 
field $\phi$ of the form:
\be \label{ey1}
\phi(x) = \wt\psi(z) + \sqrt{C\over p}\, \sum_{i=1}^{d-q-1} 
\wt\xi^i(z)\left(\sqrt 2\cos(\sqrt 2 u^i)\right)  + \cdots \, .
\ee
We have restricted $\phi$ to be even under $u^i\to - u^i$ for each 
$i$; this gives a consistent truncation of the theory at the tree
level. The dots stand for higher momentum modes which will not be 
required for our analysis. Substituting this into \refb{e1} (with 
$g$ replaced by $\wb g$) we get the action:
\ben \label{ey2}
{1\over \wb g^2}{p^2 \over p-1}\left({2\pi\over\sqrt 2}\right)^{d-q-1}
\int d^{q+1} z \bigg[ -{1\over 2} \wt\psi p^{-\half \Box_z}
\wt\psi + {1\over p+1} \wt\psi^{p+1}\nonumber \\
\qquad\qquad\qquad - C \left\{ \half\wt\xi^i
p^{-\half\Box_z} \wt\xi^i - \half\wt\psi^{p-1} \wt\xi^{i} \wt\xi^i
\right\} + \OO(\wt\xi^3) + \ldots\bigg]\, ,
\een
If we identify
\be \label{ey3}
g_q^2 = \wb g^2 \left({\sqrt 2\over2\pi}\right)^{d-q-1}\, ,
\ee
this action looks identical to the one in \refb{e13} with the
fields $\psi$, $\xi^i$ replaced by $\wt\psi$, $\wt\xi^i$ and the
identification $x_\parallel\sim z$.
In particular we can define the analogues of eqs.\refb{e14}
\be \label{ey4}
\wt\psi = 1 + {g_q \wt\sigma\over p}, \qquad \wt\xi^i = {g_q\over
\sqrt{p\, C}}
\wt\chi^i\, ,
\ee
and compute the S-matrix elements around the  vacuum $\wt\psi=1$ by
expanding \refb{ey2} in a power series in $\wt\sigma$ and $\wt\chi^i$.
Similarity of \refb{ey2} and \refb{e13} (and hence \refb{e15})
shows that the S-matrix elements computed from the action \refb{ey2}
around the $\psi=1$ background are
identical to those computed from \refb{e15} around the $\wt\psi=1$
background. In particular the S-matrix element involving 
a $\wt\chi^i$, a $\wt\chi^{j}$, and an arbitrary number of $\wt\sigma$
quanta for \refb{ey2} is identical to the S-matrix element involving
$\chi^i$, $\chi^j$ and an arbitrary number of $\sigma$ quanta in
\refb{e15}\footnote{Since the similarity of \refb{e13} and \refb{ey2}
holds only to quadratic order in $\chi$ ($\wt\chi$), we can only 
make this claim for two or less external $\chi$ ($\wt\chi$) particles.}.

On the other hand, the S-matrix elements computed from \refb{ey2} 
have direct string theory interpretation, as the action is obtained 
by compactifying a $p$-adic string theory\cite{CHEKHOV}. In particular 
the amplitude 
$\left\langle\wt\chi^i\wt\chi^{j}\wt\sigma^N\right\rangle$ is given 
in terms of correlation functions of $\wt\chi^i$, $\wt\chi^{j}$ and $N$ 
$\wt\sigma$ vertex operators on the upper half plane. The vertex operator
for 
$\wt\sigma$ is proportional to $e^{ik.Z}$, whereas that for 
$\wt\chi^i$ is given by $\sqrt 2 \cos(\sqrt 2 U^i) e^{ik.Z}$. Comparing 
this with the corresponding computation for the D-$q$-brane we see 
that the $\wt\sigma$ vertex operator is identical to the $\sigma$ 
vertex operator with $X_\parallel$ replaced by $Z$. The $\chi^i$ vertex
operator on the D-$q$-brane, 
given by $\p X_\perp^i e^{ik.X_\parallel}$, looks different from the 
$\wt\chi^i$ vertex operator even after we identify $X_\parallel$ with $Z$.
However if we note that on the boundary 
of the upper half plane the two point functions $\left\langle \sqrt 2
\cos(\sqrt 2
U^i(x_1))\sqrt 2 \cos(\sqrt 2 U^j(x_2))\right\rangle$ and 
$\left\langle \p X_\perp^i(x_1) \p X_\perp^j(x_2)\right\rangle$ 
are identical, both
being equal to $\delta_{ij} |x_1-x_2|^{-2}$, we can conclude 
that these particular amplitudes in the compactified string 
theory are indeed identical to those on the D-$q$-brane.

To summarise, we have shown that the amplitudes 
$\langle \chi^i\chi^j\sigma^N\rangle$ computed from \refb{e15} are 
identical to the corresponding amplitudes in the compactified string
theory, which in turn are identical to the corresponding ones
on the D-$q$-brane. This establishes the desired result. In presenting 
this argument we have not been careful about the overall 
normalisation factors, but the equality already established for the
amplitudes
$\langle\sigma^N\rangle$ and $\langle\chi^i\chi^j\sigma\sigma\rangle$ 
in the two theories guarantees that the overall normalisation factors 
also agree in the two theories.

This provides strong evidence that the solitonic $q$-branes of the
$p$-adic string 
theory should be identified with Dirichlet $q$-branes.

It will be interesting to systematically extend this comparison to
S-matrix elements involving more than two external $\chi^i$ states, 
and also to S-matrix elements involving higher level states. It is not 
easy to establish this in all generality, however we can
consider a subset of the massive modes on the $q$-brane and show 
the agreement between the S-matrix elements on the solitonic $q$-brane 
and D-$q$-brane with at most two of these states on the external leg. 

We start with the solitonic $q$-brane, and consider a generalisation 
of the expansion \refb{e11}: 
\be \label{ez1}
\phi(x) = F^{(d-q-1)}(x_\perp) \psi(x_\parallel) +
\sum_{r=1}^{d-q-1}{\sum_{\{i_1,\ldots i_r\}}}^{\!\!\!\!\!
\prime}\,\p_{x_\perp^{i_1}} 
\cdots \p_{x_\perp^{i_r}} F^{(d-q-1)}(x_\perp) 
\xi^{i_1\cdots i_r}(x_\parallel) + \cdots\, ,
\ee
where $\sum'$ above denotes sum over those indices $\{i_1,\cdots
i_r\}$
for which no two in the set are equal. In this case
\be \label{ez2}
\p_{x_\perp^{i_1}} \cdots
\p_{x_\perp^{i_r}}\left(F^{(d-q-1)}(x_\perp)\right)^p = p^r \left(
F^{(d-q-1)}(x_\perp)\right)^{p-1}\p_{x_\perp^{i_1}} \cdots
\p_{x_\perp^{i_r}} F^{(d-q-1)}(x_\perp)\, .
\ee
Substituting \refb{ez1} into the equation of motion \refb{e1b}, and using
eq.\refb{ez2} we get
\ben \label{ez3}
p^{-\half\Box_\parallel} \psi &=& \psi^p + \OO(\xi^2)\, \nonumber \\
p^{-\half\Box_\parallel} \xi^{i_1\cdots i_r} &=& p^{1-r}
\psi^{p-1}\xi^{i_1\cdots i_r} + \OO(\xi^2)\, .
\een
The action involving these fields is given by
\ben \label{ez4}
S_q(\psi,\xi) 
&=& {1\over g_q^2} {p^2 \over p-1} \int d^{q+1}x_\parallel \Bigg[ 
-{1\over 2} \psi p^{-\half\Box_\parallel}
\psi + {1\over p+1} \psi^{p+1} \nonumber \\
&& - \sum_{r=1}^{d-q-1}{\sum_{\{i_1,\ldots i_r\}}}^{\!\!\!\!\!\prime}\, 
C_r \left\{
{1\over 2}\xi^{i_1\cdots i_r} p^{-\half \Box_\parallel + r -1}
\xi^{i_1\cdots i_r} - {1\over 2} \psi^{p-1} \xi^{i_1\cdots i_r}
\xi^{i_1\cdots i_r}\right\} + \OO(\xi^3)\Bigg]\, .
\nonumber \\
\een
$C_r$ is a normalisation constant which can be absorbed into the
definition of $\xi^{i_1\cdots i_r}$.
{}From this we see that the mass-shell constraint for 
$\xi^{i_1\cdots i_r}$, in the $\psi=1$ background, is
\be \label{ez5}
k^2 = 2(1 - r)\, .
\ee

In order to compare this with the world-volume theory on the D-$q$-brane,
we need to first identify the vertex operator
corresponding to the mode $\xi^{i_1\cdots i_r}$. We
take this to be
\be \label{ez6}
V_{i_1\cdots i_r} = \p X_\perp^{i_1} \cdots \p X_\perp^{i_r} 
e^{ik.X_\parallel}\, .
\ee
This describes a physical state satisfying the same mass shell constraint
as eq.\refb{ez5} as long as the indices in the set 
$\{i_1,\ldots i_r\}$ are all different.

We shall now compare the S-matrix elements computed from the action
\refb{ez4} with two or less external $\xi$-legs to that computed 
directly in the D-$q$-brane. For two external tachyons and two external
$\xi$ we can do this
explicitly and verify that it agrees with the corresponding 
computation on the D-$q$-brane. The computation is identical to the one 
discussed earlier. For arbitrary number of external tachyon legs, one
can generalise the argument given for external $\chi^i$-legs. 
The key ingredient of this argument is that the two point function 
of $\p X_\perp^{i_1} \cdots \p X_\perp^{i_r} e^{ik.X_\parallel}$ and 
$\p X_\perp^{i_1'} \cdots \p X_\perp^{i_r'} e^{ik'.X_\parallel}$ for
the D-$q$-brane is identical to that between the vertex operators 
$(\sqrt 2)^r \cos(\sqrt 2 U^{i_1}) \ldots\cos(\sqrt 2 U^{i_r}) e^{ik.Z}$
and $(\sqrt 2)^r \cos(\sqrt{2}U^{i_1'})\ldots \cos(\sqrt 2 U^{i_r'})
e^{ik'.Z}$ of the compactified string theory.

\sectiono{Comments} \label{s5}

\begin{itemize}
\item {}We have shown that one can get a consistent truncation of the 
world-volume theory on a D-$q$-brane in $p$-adic string theory by 
keeping only the tachyonic mode. Thus by examining the tree level 
tachyon amplitudes in the world-volume theory we shall not discover the 
existence of the other modes. This suggests that there may be other
(massless and massive)
modes living on the world-volume of the space-filling D-$(d-1)$-brane 
as well, inspite of the fact that there are no poles in the tachyon
S-matrix elements 
corresponding to these states. Indeed, ref.\cite{MAZA} attempted to 
generalise the $p$-adic string amplitudes to external vector states. 
If these modes are present they will give rise to new degrees of 
freedom on the solitonic $q$-brane, and will have to be taken into 
account in comparing the world-volume theory on the D-$q$-brane with 
that on the solitonic $q$-brane.

\item It will also be of interest to compute the tension of a Dirichlet
$q$-brane in the $p$-adic string theory independently, and compare with
eq.\refb{e4} describing the tension of a solitonic $q$-brane. This will
require careful analysis of the cylinder amplitude, and a proper
understanding of the closed string sector of the theory.

\item It has been shown in ref.\cite{padic} that it is possible to assign
Chan-Paton factors to the open string states of a $p$-padic string theory.
This shows the existence of multiple D-$(d-1)$-branes. Furthermore if
there are massless gauge fields in the spectrum of open strings in the
$p$-adic string theory, and if there is a T-duality
transformation relating the
D-$(d-1)$-brane to D-$q$-brane, then by
switching on Wilson lines corresponding to the gauge fields followed by a
T-duality transformation, we can produce static configuration of
D-$q$-branes separated in space. It will be interesting to examine if the
equation of motion \refb{e1b} admit such solutions. Ideas developed
in ref.\cite{0003160} may be useful in this context.

\item It is natural to ask if this analysis has any relevance to the 
ordinary bosonic string theory. Firstly, we would like to point out
that even if the tachyon potential in the $p$-adic string theory 
is totally unrelated to that in the ordinary bosonic string theory, it 
can be regarded as a toy model which nicely illustrates the features 
expected of the full bosonic string field theory action. Besides, there 
is evidence of close relationship between tachyon amplitudes in the 
$p$-adic and ordinary bosonic string
theory\cite{FRWI,ZABRODIN}. 
Thus one might hope that the full tachyon effective action in bosonic 
string field theory is related in some way to the tachyon effective 
action in $p$-adic string theory. 

In this direction, we cannot resist 
the temptation to point out some apparent similarities between the 
equation of motion \refb{e1b} and that in the open bosonic string 
field theory\cite{WITTENBSFT}. To lowest order in the level truncation 
scheme\cite{KS}, the tachyon equation of motion in open bosonic string 
field theory may be written as\cite{0003031}
\be \label{ezz1}
\left[\left(\alpha'\Box + 1\right) e^{-{c\,\alpha'}\Box} -
2\right] \phi = \wb g \phi^2\, ,
\ee
where $c=\ln(3^3/4^2)$, $\wb g$ is open string coupling constant after
suitable normalisation, and $\phi$ is related to the original tachyon
field $T$ by a field redefinition $T=e^{-c\,\alpha'\Box/2}\phi + \wb
g^{-1}$, 
so that $\phi=0$ is the vacuum without any D-brane, and 
$\phi=-1/\wb g$ denotes the D-brane. If we drop the first and 
the third terms on the left hand side of eq.\refb{ezz1} by hand, then 
this equation, after suitable rescaling of $x$ and $\phi$, reduces to
eq.\refb{e1b} for $p=2$. Of course there is no justification for dropping
these terms, so we shall not pursue this matter any further; but it is
not inconceivable that some exact relation between bosonic string field
theory and $p$-adic string theory will be discovered in the future.

\end{itemize}

\noindent {\bf Acknowledgement}: We wish to thank S.\ Mukhi for useful
discussions.

\end{document}